\documentclass[twoside,twocolumn,9pt]{article}
\usepackage{extsizes}
\usepackage[super,sort&compress,comma]{natbib} 
\usepackage[left=1.5cm, right=1.5cm, top=1.785cm, bottom=2.0cm]{geometry}
\usepackage{balance}
\usepackage{times,mathptmx}
\usepackage{sectsty}
\usepackage{graphicx} 
\usepackage{lastpage}
\usepackage[format=plain,justification=justified,singlelinecheck=false,font={stretch=1.125,small,sf},labelfont=bf,labelsep=space]{caption}
\usepackage{float}
\usepackage{fancyhdr}
\usepackage{fnpos}
\usepackage[english]{babel}
\addto{\captionsenglish}{%
	
}
\usepackage{array}
\usepackage{charter}
\usepackage[T1]{fontenc}
\usepackage{setspace}
\usepackage[compact]{titlesec}
\usepackage{hyperref}
\usepackage{multirow}
\usepackage{amsmath}
\usepackage{float}
\restylefloat{table}
\usepackage{subfigure}
\usepackage{booktabs}
\usepackage[table,xcdraw]{xcolor}
\definecolor{cream}{RGB}{222,217,201}
\usepackage{adjustbox}
\usepackage{lipsum}

\usepackage{amsmath}
\usepackage{amssymb}
\usepackage{graphicx}
\usepackage{xcolor}
\usepackage{blindtext}
\usepackage{multirow}
\usepackage{chngcntr}
\usepackage{subfigure}

	\def\k{{\bf k}}
	\def\etal{{\it et al. }}

\begin{document}

\pagestyle{fancy}
\thispagestyle{plain}

\makeFNbottom
\makeatletter
\renewcommand\LARGE{\@setfontsize\LARGE{15pt}{17}}
\renewcommand\Large{\@setfontsize\Large{12pt}{14}}
\renewcommand\large{\@setfontsize\large{10pt}{12}}
\renewcommand\footnotesize{\@setfontsize\footnotesize{7pt}{10}}
\makeatother

\renewcommand{\thefootnote}{\fnsymbol{footnote}}
\renewcommand\footnoterule{\vspace*{1pt}%
	\color{cream}\hrule width 3.5in height 0.4pt \color{black}\vspace*{5pt}} 
\setcounter{secnumdepth}{5}

\makeatletter 
\renewcommand\@biblabel[1]{#1}            
\renewcommand\@makefntext[1]%
{\noindent\makebox[0pt][r]{\@thefnmark\,}#1}
\makeatother 
\renewcommand{\figurename}{\small{Fig.}~}
\sectionfont{\sffamily\Large}
\subsectionfont{\normalsize}
\subsubsectionfont{\bf}
\setstretch{1.125} 
\setlength{\skip\footins}{0.8cm}
\setlength{\footnotesep}{0.25cm}
\setlength{\jot}{10pt}
\titlespacing*{\section}{0pt}{4pt}{4pt}
\titlespacing*{\subsection}{0pt}{15pt}{1pt}

\fancyfoot{}
\fancyfoot[RO]{\footnotesize{\sffamily{1--\pageref{LastPage} ~\textbar  \hspace{2pt}\thepage}}}
\fancyfoot[LE]{\footnotesize{\sffamily{\thepage~\textbar\hspace{3.45cm} 1--\pageref{LastPage}}}}
\fancyhead{}
\renewcommand{\headrulewidth}{0pt} 
\renewcommand{\footrulewidth}{0pt}
\setlength{\arrayrulewidth}{1pt}
\setlength{\columnsep}{6.5mm}
\setlength\bibsep{1pt}

\makeatletter 
\newlength{\figrulesep} 
\setlength{\figrulesep}{0.5\textfloatsep} 

\newcommand{\topfigrule}{\vspace*{-1pt}%
	\noindent{\color{cream}\rule[-\figrulesep]{\columnwidth}{1.5pt}} }

\newcommand{\botfigrule}{\vspace*{-2pt}%
	\noindent{\color{cream}\rule[\figrulesep]{\columnwidth}{1.5pt}} }

\newcommand{\dblfigrule}{\vspace*{-1pt}%
	\noindent{\color{cream}\rule[-\figrulesep]{\textwidth}{1.5pt}} }

\makeatother

\twocolumn[
\begin{@twocolumnfalse}
	
	\sffamily
	\begin{tabular}{m{4.5cm} p{13.5cm} }
		
		\noindent{\textbf{DOI:}} & 
		\noindent\LARGE{\textbf{Electronic structural critique of interesting thermal and optical properties of C$_{17}$Ge germagraphene.}} \\
		\vspace{0.3cm} \\  
		
		& \noindent\large{Sujoy Datta,\textit{$^{a,b}$}  Debnarayan Jana\textit{$^{a}$} } \\
		\\
		
		\noindent{\textbf{DATE: \date{\today} }} & 
		\noindent\normalsize{
In this communication, we report a theoretical attempt to understand the involvement of electronic structure in determination of optical and thermal properties of C$_{17}$Ge germagraphene, a buckled two dimensional material. The structure is found to be a direct bandgap semiconductor with low carrier effective mass. Our study has revealed the effect of spin-orbit coupling on the band structure and in appearance of spin Hall current in the material. A selectively high blue to ultraviolet light absorption and a refractive index comparable to flint glass open up the possible applicability of this material for optical devices. From electronic structural point of view, we investigate the reason behind its moderately high Seebeck coefficient and power factor comparable to traditional thermoelectric materials. Besides its narrow bandgap, relatively smaller work function of C$_{17}$Ge ($4.361 ~eV$) than graphene ($4.390 ~eV$) and germanene ($4.682 ~eV$) assures more easily removal of electron from the surface. This material is turned out to be an excellent alternative for futuristic semiconductor application from optical to thermal device regime.
}

\end{tabular}
\end{@twocolumnfalse} \vspace{0.6cm}
]

\renewcommand*\rmdefault{bch}\normalfont\upshape
\rmfamily
\section*{}

\footnotetext{\dag corresponding author: djphy@caluniv.ac.in}
\footnotetext{\textit{$^{a}$Department of Physics, University of Calcutta, Kolkata 700009, India.}}
\footnotetext{\textit{$^{b}$Department of Physics, Lady Brabourne College, Kolkata 700017, India. }}

\vspace{-1cm}

\section{Introduction}
{\par}The discovery of graphene \cite{graphene} and related compounds revolutionized modern semiconductor industry. Subsequently, other 2D materials like silicene \cite{silicine}, black-phosphorene \cite{phosphorene}, and borophene \cite{borophene} were synthesized experimentally. 
 
{\par}Though graphene exhibits extraordinary thermal, mechanical and electrical properties, its zero band-gap imposes a severe constraint on its practical applicability. As a result, opening up the band-gap of graphene and other planar zero-gap materials has been considered to be the  top-most priority in semiconductor engineering. It has been seen earlier that introducing defects in graphene or graphene-nanotubes has a large influence on their electronic properties \cite{dope1, dope3, das2016, nanotube_dope, nanotube_dope2, nanotube_sujoy}. 

{\par}Experimental work on such materials have been prolific. The aim of this communication is to use the latest theoretical approaches, together with what we shall argue to be  better modifications, and explain these experimental results. Dependable theoretical predictions will then provide the experimentalist a much better handle in choosing their desired materials out of a plethora of possibilities.

{\par}Silicon and Germanium being the same group IV material as Carbon, are always the first choice for doping graphene. Several theoretical studies on silicon doped graphene, i.e., siligraphene \cite{siligraphene1, siligraphene2, siligraphene3} have been carried out by varying the relative concentrations of Silicon and Carbon. Siligraphene with $Si:C=1:7$ showed superior sunlight absorbance \cite{siligraphene1}. Wang \etal found theoretically that siligraphene had wonderful Li-ion storage capacity \cite{wang2018}. Dong \etal have reported that the stable siligraphene structure (SiC$_7$) exhibited a direct band-gap and superior light absorbance making it a promising donor material for optical-devices \cite{siligraphene1}.

{\par}Theoretically, Ge doping was seen to tune the band-gap \cite{germagraphene-bg, germagraphene-bg2} but these studies never explored the situation of buckled structure.  As Ge is much heavier atom than C, doping with Ge is supposed to result in off-the-plain buckling and it is never easy to dope graphene by such a heavier Ge atom. Very recently Tripathi \etal successfully implanted Ge on graphene \cite{expt-germagraphene} and found a buckled structure when a Ge atom replaces a C atom. However, when Ge replaces a C-C bond, it can be accommodated in the plane. Hu \etal found buckled C$_{17}$Ge to be stable and reported lithium ion absorption in this buckled structure \cite{germagraphene-lithium}. However, for functionalization of the germagraphene, electronic structural study should be done more extensively. Here, in this systematic study, we try to explore the electronic, optical and thermal properties of the germagraphene structure.


\section{Computational Details}
{\par}The basic calculations are carried out in plane wave based techniques used in the Quantum Espresso (QE) code \cite{QE1, QE2}. Slab geometry for two dimensional system is simulated by introducing 12 \AA ~vacuum on  either side of the sheet. The structure is energetically optimized first using variable cell structural relaxation technique. Projected augmented wave (PAW) basis of the Quantum Espresso (QE) is utilised using Perdew-Burke-Ernzerhof (PBE) exchange-correlation potentials\cite{PBE,PBE1}. Charge-densities and energies for each calculation were converged to 10$^{-7}$ Ry. with the maximum force of 0.001 Ry./atom.  $6\times 6\times 2$ k-point meshes are used and force convergence and pressure thresholds are set as 0.0001 Ry./au and 0.0000 Kbar, respectively. The self consistent field (SCF) calculation on the relaxed structure is done using $8\times 8\times 1$ dense k-point mesh for PBE and $4\times 4\times 1$ k and q point meshes for HSE calculation. 

{\par} For calculation of spin Hall conductivity (SHC) and transport properties we have used Wannier90 \cite{wannier90_pizzi2019} package for constructing the maximally localized Wannier functions (MLWF). For construction of MLWF relatively coarse k-ponts as applied in SCF calculation is enough. However, for SHC and transport calculations much dense k-points is needed \cite{PhysRevLett.94.226601}. That is why utilising the Wannier functional way is always beneficial \cite{qiao2018}. A fine sample of $400 \times 400 \times1$ and $100 \times 100 \times1$ k-points are used for SHC and thermal property calculations, respectively. Wannier interpolation technique is also applied for the HSE band plotting.

{\par} BoltzWann module of Wannier90 is a powerful tool for calculating transport properties of material at moderate computational cost \cite{pizzi2014}. The high accuracy in the Brillouin zone integrals can be achieved through Wannierization. In thermal properties calculations we have used adaptive interband smearing and constant relaxation-time of $10fs$.

{\par}For optical property predictions, we calculate the complex dielectric tensor using random phase approximation (RPA).
\begin{align}
\epsilon_{\alpha\beta}(\omega)&= 1+\frac{4 \pi e^2}{\Omega N_{\textbf{k}} m^2}\sum\limits_{n,n'}\sum\limits_{\textbf{k}}
\frac{\langle u_{\textbf{k},n'}\vert\hat{\textbf{p}}_{\alpha}\vert u_{\textbf{k},n}\rangle 
\langle u_{\textbf{k},n}\vert\hat{\textbf{p}}_{\beta}^{\dagger} \vert u_{\textbf{k},n'}\rangle}
 {(E_{\textbf{k},n'}-E_{\textbf{k},n})^2} \nonumber\\
&\left[\frac{f(E_{\textbf{k},n})}{E_{\textbf{k},n'}-E_{\textbf{k},n}+(\hbar\omega+i\hbar\Gamma)} +
\frac{f(E_{\textbf{k},n})}{E_{\textbf{k},n'}-E_{\textbf{k},n}-(\hbar\omega+i\hbar\Gamma)}\right] 
\end{align}

{\par}Here, $\Gamma$ is the inter-smearing term tending to zero. Since no excited-state can have infinite lifetime, we have introduced small positive $\Gamma$ in order to produce an intrinsic broadening to all excited states. The imaginary part of the dielectric function $\epsilon^{(i)}_{\alpha\beta}$ has been calculated first and the real part $\epsilon^{(r)}_{\alpha\beta}$ is found using the Kramers-Kronig relation.
Optical-conductivity, refractive index and absorption-coefficients were calculated using real and imaginary parts of dielectric functions.\cite{wooten}
\begin{align}
&\text{Dielectric tensor:  } \epsilon_{\alpha\beta}=\epsilon_{\alpha\beta}^{(r)}+ i \epsilon^{(i)}_{\alpha\beta} \\ 
&\text{Optical Conductivity:  } Re [\sigma_{\alpha\beta} (\omega)]= \frac{\omega}{4\pi}\epsilon^{(i)}_{\alpha\beta}(\omega) \\
&\text{Complex Refractive Index:  } \mu_{\alpha\alpha}=n^+_{\alpha\alpha}+i n^-_{\alpha\alpha}\\ 
&\text{Absorption Coefficient: } A_{\alpha\alpha}(\omega)=\frac{2\omega n^-_{\alpha\alpha}(\omega)}{c} \label{abs}\\ 
& \text{where, } n^{\pm}_{\alpha\alpha}(\omega)= \sqrt{\frac{|\epsilon_{\alpha\alpha}(\omega)| \pm \epsilon^{(r)}_{\alpha\alpha}(\omega)}{2}} \nonumber
\end{align}

\section{Results and discussion}


\subsection{Structural Properties: }
\begin{figure}[t!]
	\centering	
	{\includegraphics[scale=0.45]{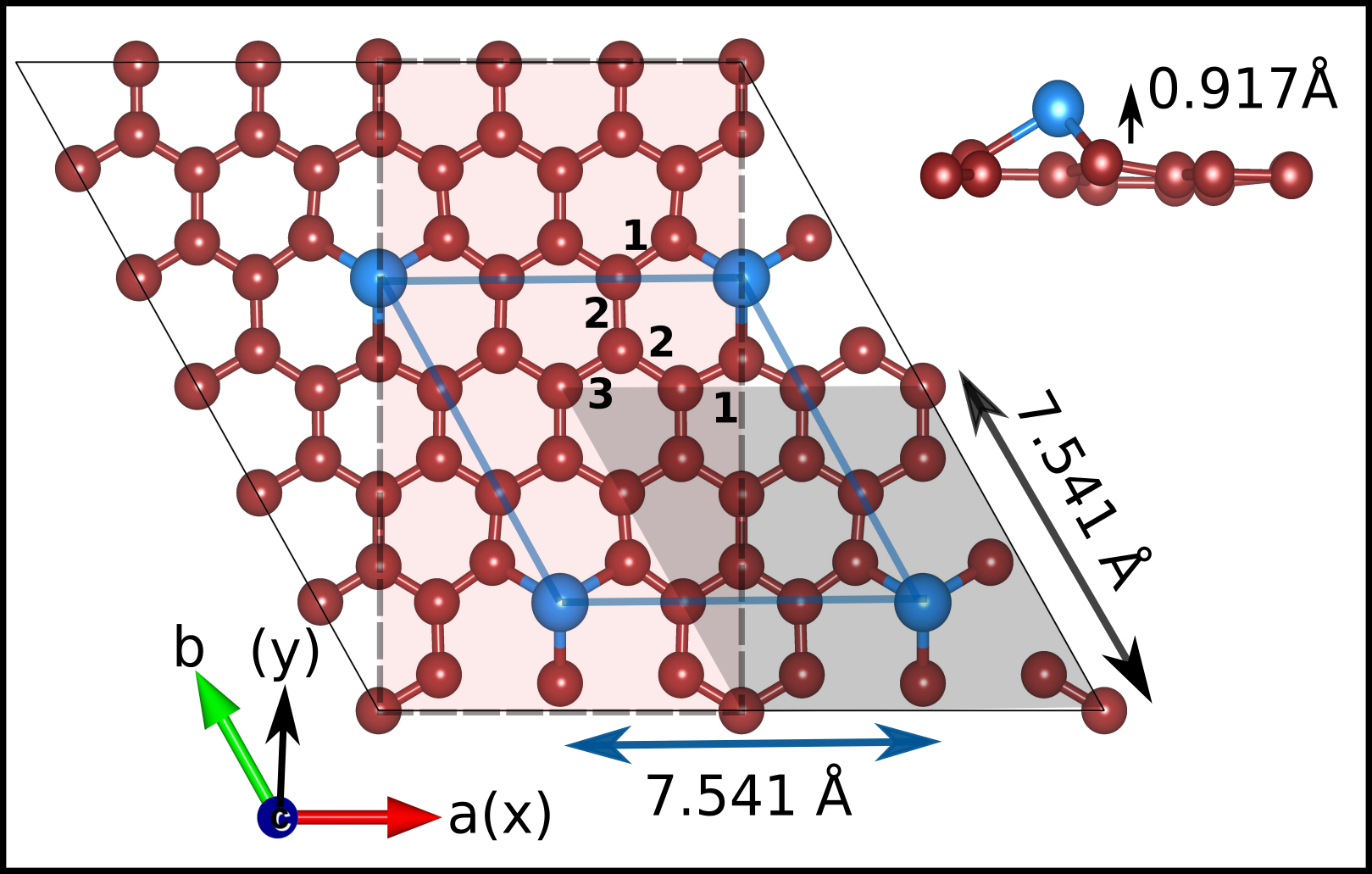}}
	\caption{2$\times$2 supercell structure of C$_{17}$Ge. a and b are the primitive vectors. The shaded region in grey shows the primitive unit cell and the shaded region in pink is used for elastic properties calculation. Optimized bond lengths labelled by 1, 2, 3 are 1.411\AA, 1.466\AA, 1.450 \AA, while the C-Ge bond-length is 1.863\AA. C-Ge-C, C-C-Ge and C-C-C bond angles are 97.834$^o$, 115.987$^o$ and 122.968$^o$.}
	\label{fig1}
\end{figure}

{\par}A 3$\times$3 supercell of graphene (P6/mmm symmetry group) containing 18 C atoms was built. Then a C atom was replaced by Ge and the structure was relaxed. The unit cell of the C$_{17}$Ge structure is shown as shaded region in Fig. \ref{fig1}.

{\par}The geometrically relaxed structure shows buckling. Our calculations complement the experimental finding of buckled germagraphene structure \cite{expt-germagraphene} as well as the theoretical result of Hu. \etal \cite{germagraphene-lithium} who also showed the dynamical stability of the structure. Being a heavier atom, Ge tends to distort the planar graphene structure more than that of siligraphene structures \cite{siligraphene1}. The in-plane lattice constant is 7.541 \AA ~and the Ge atom at 0.917 \AA ~off-the-plane. The separation between layers is kept at 12 \AA ~to nullify the intra-layer interaction, prerequisite for 2D calculations. We can clearly visualize a rhombic formation of Ge atoms connected by blue lines in Fig.\ref{fig1}. C-Ge bond length is 1.863 \AA, whereas, the C-C bonds denoted by 1, 2, 3 have lengths of  1.411, 1.466, 1.450 \AA ~ respectively. So, all the hexagonal rings with only carbon atoms are not regular-hexagons after Ge doping. C-Ge-C, C-C-Ge and C-C-C bond angles are 97.834$^o$, 115.987$^o$ and 122.968$^o$, respectively. The buckled structure suggests a deviation from the pure sp$^2$ hybridization which is a signature of planar structure.

\vskip 1cm
\textit{\textbf{ Elastic constants: }}

\begin{figure}[b!]
	\centering
	\includegraphics[scale=0.3]{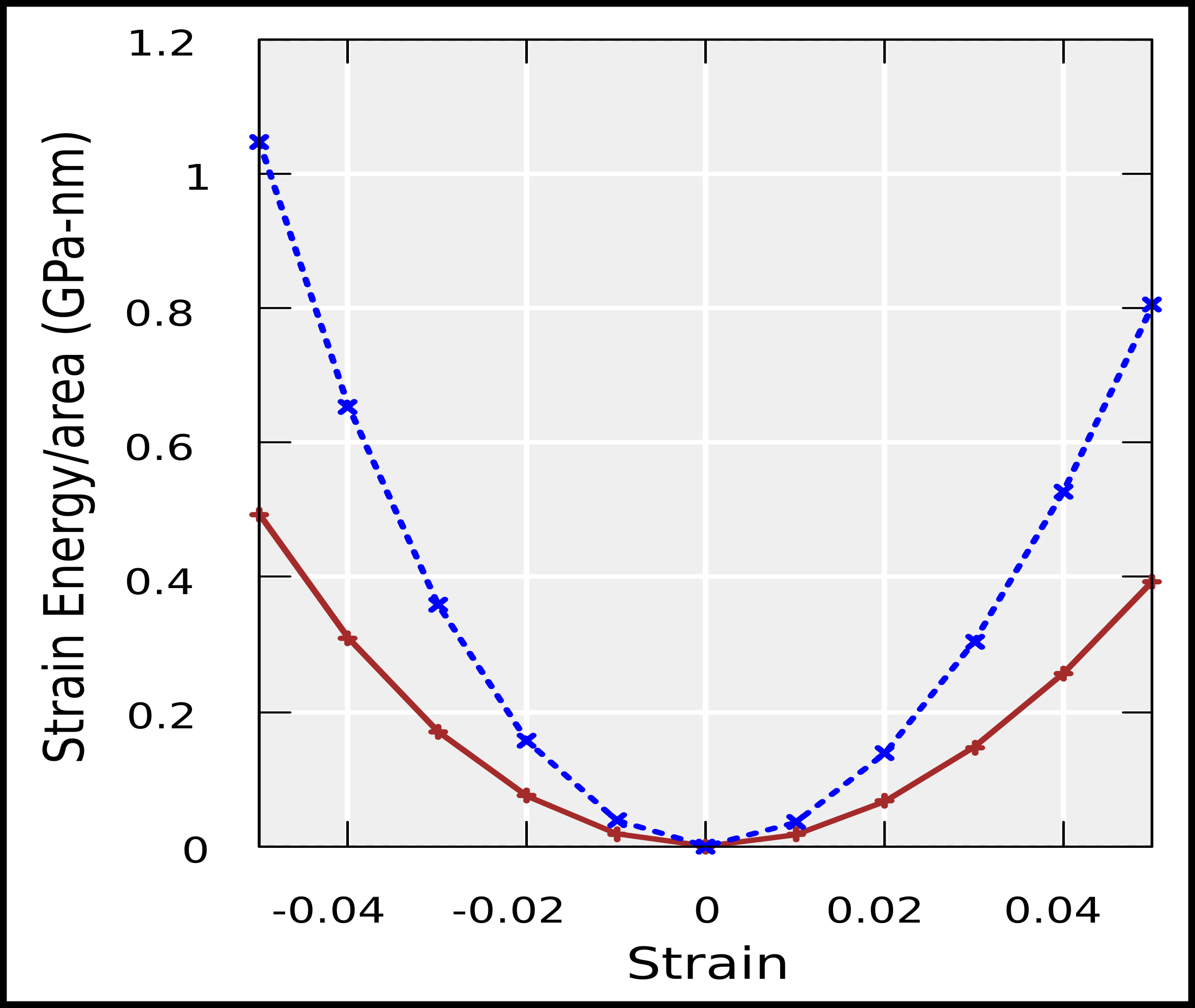}
	\caption{Strain energy with respect to uniaxial (blue-dashed) and biaxial (brown-solid) strain.}
	\label{elastic}
\end{figure}

\begin{figure*}[t!]
	\centering
	\includegraphics[scale=0.55]{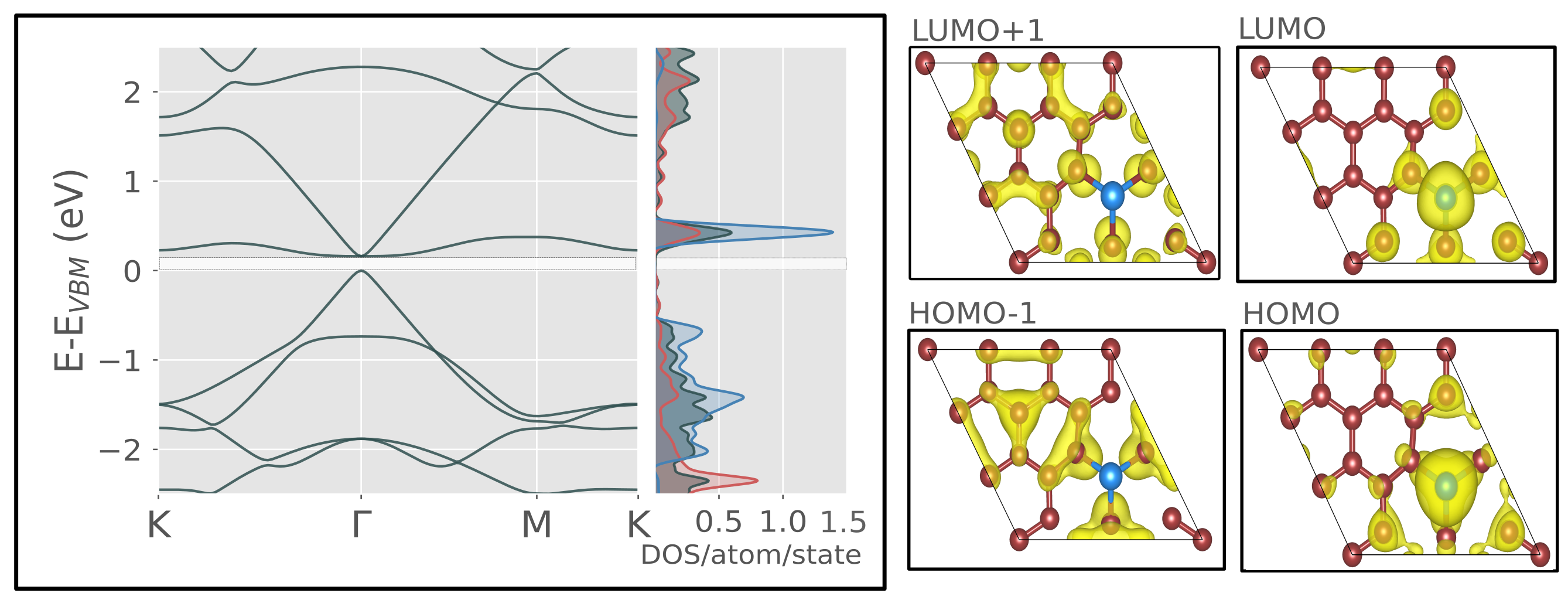}
	\caption{(left)The band structures and density of states for C$_{17}$Ge using PBE. The p$_z$ projected partial DOS of Ge and C are shown in blue and red, whereas, the total DOS is in grey. The $\Gamma - \Gamma$ direct band gap is $0.166 ~eV$ between HOMO originating from C (in red) and LUMO from Ge (in blue). (right) Charge density plots for the states near Fermi energy.}  
	\label{fig2}
\end{figure*}

{\par} In Fig. \ref{fig1} the shaded region in pink (dashed line) is used as the cell for the energy of state (EOS) calculation. Uniaxial strain is produced along either in direction (x) or in direction (y). For biaxial stress condition, same percentage of strain is produced along both direction. The total energy is fitted using the equation \cite{liu2018}:
\begin{equation}
u =\frac{1}{2}C_{11} \delta_x^2 + \frac{1}{2}C_{22} \delta_y^2 + C_{12} \delta_x \delta_y + 2 C_{66} \delta_{xy}^2
\end{equation}
where, $u$ is the the strain energy per unit area of two dimensional structure. In standard Vigot notation $\delta_x$ and $\delta_y$ are the uniaxial in plane strain and $\delta_{xy}$ is the sheering strain. As the structure is symmetrical along x and y, so, $C_{11}=C_{22}$.

{\par}In Fig.\ref{elastic} the EOS plot is depicted. The elastic constants are found from the fitting curve as $C_{11}= 354.607$ GPa-nm and $C_{12}= 15.272$ GPa-nm. These satisfy the Born criteria for mechanical stability of 2D materials, $C_{11} \times C_{22} \rangle C_{12}^2$. The in plane Young's modulus as derived from the formula $Y=\frac{C_{11}^2-C_{12}^2}{C_{11}}$ is 353.949 GPa-nm which is higher than that of graphene (335 GPa-nm) \cite{csahin2009} and about three times of MoS$_2$ (123 GPa-nm) \cite{yue2012}. The Poission's ratio defined as $\upsilon=\frac{C_{12}}{C_{11}}$ is 0.043.

\subsection{Electronic Properties:}
{\par}The band structure and densities of states (DOS) for C$_{17}$Ge are plotted in Fig.\ref{fig2} . The valence band maxima (VBM) is taken as the reference zero of energy axis. From band structure plot it is evident that C$_{17}$Ge is a direct band-gap semiconductor. The estimated band-gap using PBE is $0.166 ~eV$ at $\Gamma$ point.

{\par}Near the Fermi energy (E$_F$), there is one band approaching from below, the highest occupied (HO) band. In conduction band region, there are two lower lying bands, one is almost flat, the lowest unoccupied (LU) and another is sharp dipping (LU+1). In the vicinity of $\Gamma$, HO and LU$+1$ bands show almost linear dispersion if moved along $\Gamma (0,0,0)$ to $K (1/3,1/3,0)$ as well as along $\Gamma (0,0,0)$ to $M (1/2,0,0)$. So, similar E-k relation is found for these two bands, both along $k_{b1}$ and $k_{b2}$, where $b1$ and $b2$ are the reciprocal lattice vectors in $k_xk_y$ plane for hexagonal system. Such dispersion isotropy is also present in LU but with much smaller curvature which will affect in the mobility of carriers. However, the dispersive nature of HO$-1$ and LU$+1$ bands along $\Gamma -K$ and $\Gamma - M$ paths are not the same. A HO and HO$-1$ crossover is seen at about the midway of $\Gamma - M$. At $\Gamma$ there is a band degeneracy in conduction band as LU and LU$+1$ bands overlap.

{\par}The total density of state (DOS/atom/state) is plotted in grey. Approaching E$_F$ from the occupied states below, the total DOS (TDOS) shows a gradual fall, whereas, there is a sharp peak at the lowest unoccupied state. To understand the origin of the bands, we plot the p$_z$ projected DOS (pDOS) of C and Ge in red and blue, respectively. The mixing of Ge and C states are evident from pDOS plots. The sharp pDOS peaks of both Ge and C just near CBM suggests the dominant existence of Ge-p$_z$ state in LU. The flatness of LU has given rise to this sharp peak, i.e., higher density of states. In valence band (VB) region, the Ge energy states are more available near the E$_F$ than at lower energy. Except the sharp peak in LU, the C pDOS rises gradually in both higher and lower energy direction with respect to E$_F$. So, leaving the peak, this nature is a reflection of linear dispersive systems, small DOS around E$_F$. This nature of carbon pDOS and the dominant existence of Ge states in the vicinity of Fermi energy, both in CB and VB confirm the role of Ge in opening the gap between the HO and LU.   

{\par}The logical conclusion on the origin of the bands near Fermi energy finds a stronger base from the charge density plots in Fig. \ref{fig2}. We have plotted the charge densities for HO, LU as well as HO$-1$ and LU$+1$ states. These plots are helpful to identify active sites in C$_{17}$Ge. The HO and LU states dominated by Ge-$p_z$ states whereas, HO$-1$ and LU$+1$ states are composed of C-Ge bonding states with smaller contribution from Ge. As none of the HO, HO$-1$, LU, LU$+1$ sates are localized, the efficiency of photo-absorption is expected. The prevalent existence of Ge-$4p$ charge densities in HO and LU reassert the role of Ge in band-gap opening.  

\vskip 1cm
\noindent
\textit{\textbf{Carrier Mobility: }}
{\par}Due to the degeneracy of bands at CBM, one flat and a sharp band, it would be interesting to calculate the carrier mobility. Mobility is the defining factor of the dynamical activity of the charge careers in any semiconductor. Higher mobility means speedy charge transfer and low energy loss, yielding better performance and durability of an electronic device.

{\par}Mobility is inversely proportional to the effective mass (m$^{\ast}$) of carrier hole or electron. From the band-structure, $m^{\ast}$ can be calculated as $m^{\ast}={\hbar^2}/\frac{d^{2}E}{dk^{2}}$ near the band-edges, for electron at CBM and at VBM for hole.

{\par}The calculated hole effective mass is $m^{\ast}_h=-0.00292 m_0$ where $m_0$ is electron rest mass. The electron effective mass has two different values due to the band degeneracy at CBM (at $\Gamma$). The electron effective mass originating from LU, $m^{\ast}_e(L)= 0.59528 m_0$ is less than $\frac{1}{100} th$ that of $m^{\ast}$ for the LU$+1$ band, $m^{\ast}_e(L+1)=0.00291 m_0$. Such low light-electron and hole effective mass indicate very high carrier mobility in germagraphene, which is much higher than many 2D materials and reported so far\cite{miro2014}.

{\par}As we find C$_{17}$Ge as a narrow gap semiconductor, so, its applicability should be compared with the range of II-VI narrow gap semiconductors. For example, both the light-electron and hole effective masses of this germagraphene are almost $\frac{1}{10}th$ to those of HgCdTe ($m^{\ast}\approx 0.071 m_0$) \cite{rogalski2005}. So, C$_{17}$Ge readily can be considered as a better option for the optoelectronic devices like infrared detector, photovoltaic devices etc., those  now built by alloying semimetals with II-VI semiconductors \cite{baker2007}.  

\vskip 1cm
\noindent
\textit{\textbf{Effect of Intrinsic Spin-Orbit Coupling: }}
{\par}For those systems where space inversion symmetry holds, time reversal symmetry would produce double degeneracy of the bands. Graphene, like many other two-dimensional (2D) materials, preserves inversion symmetry. So, even in the presence of intrinsic spin-orbit coupling (SOC), the Bloch states remain doubly spin degenerate at $K$ where the Dirac point originally formed, however, a gap opens up \cite{han2014}. Application of transverse external electric field can lift the spin degeneracy by the Rashba effect. Adatoms having stronger spin-orbit coupling can be doped to open up the band-gap of graphene \cite{weeks2011}. The system we are dealing here can be otherwise viewed as a periodically doped graphene and having a band gap as well, even in the absence of SOC. So, in presence of intrinsic SOC it is interesting to explore the band structure of C$_{17}$Ge.

\begin{figure}[h!]
	\centering
	\includegraphics[scale=0.3]{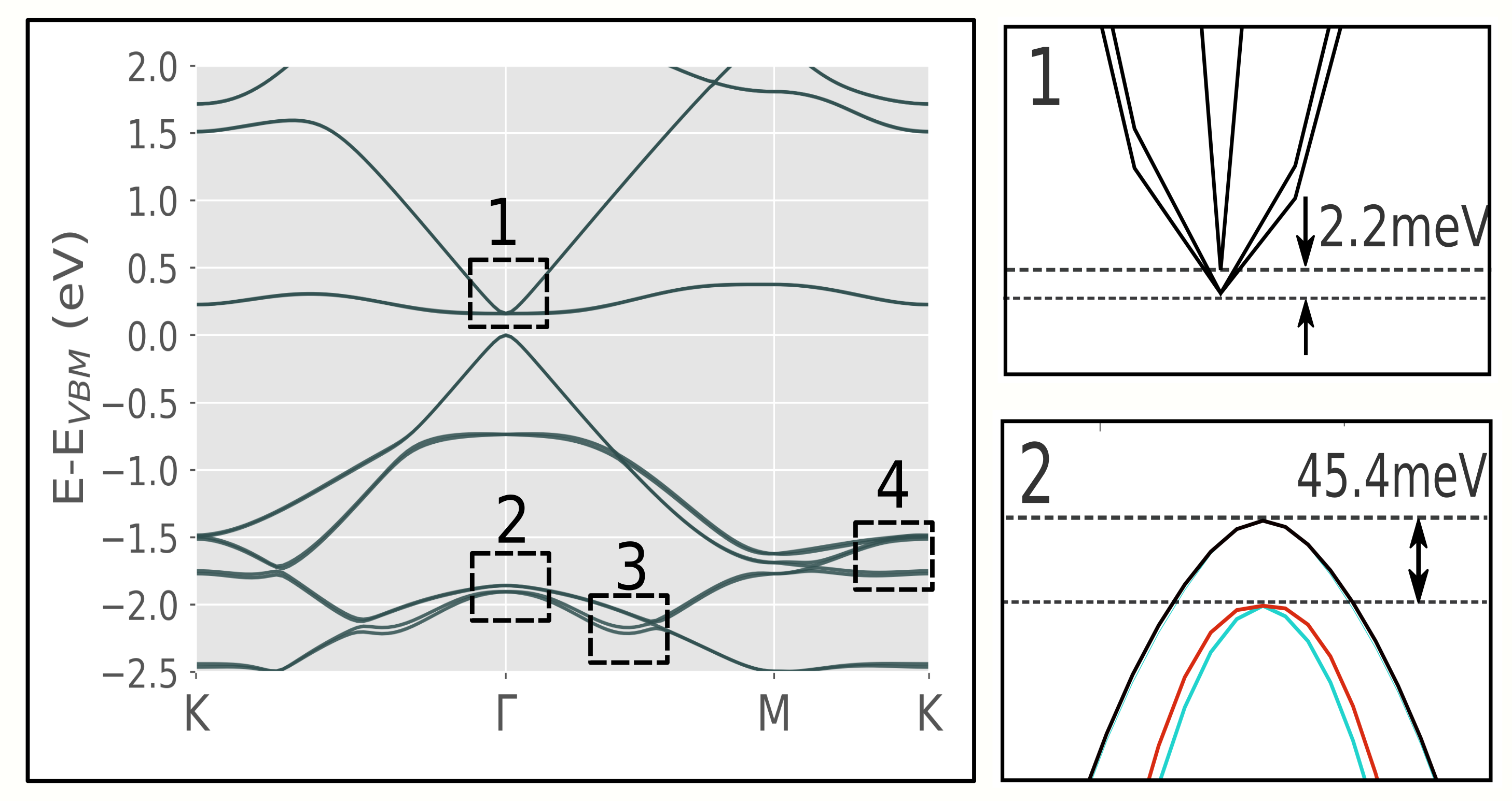}
		\caption{Band structure of C$_{17}$Ge with spin-orbit coupling (SOC). Different regions of interest are enlarged.}
	\label{fig3}
\end{figure}

{\par}Band structure of C$_{17}$Ge in presence of intrinsic SOC is shown in Fig.\ref{fig3}. Though the plot looks almost same as that without SOC (Fig.\ref{fig2}), a close look at several region reveals the difference as represented in the insets. The $\Gamma$ point, being the centre of reciprocal space, is the most symmetrical point preserving both time reversal and inversion symmetries. So, no spin-orbit spiting is expected at $\Gamma$ and the regions denoted by 1 and 2 endorse that. However, the band degeneracy of LU and LU$+1$ is now removed at $\Gamma$ point by a small value of $2.2 ~meV$ (Fig.\ref{fig3}-1). This has an effect on the effective mass of electron as removal of band degeneracy now confirms this value $m^{\ast}_e(L)= 0.0302 m_0$. 

{\par}The HO$-1$ band has considerable C and Ge state mixing and as Ge atom is much heavier than C, spin-orbit coupling effect should be more evident in that band. In the vicinity of $\Gamma$ point, the splitting is observable and at the Brilouin zone edge $K (1/3,1/3,0)$ the splitting is as large as $21.8 ~meV$ (Fig.\ref{fig3}-4). The degenerate bands in occupied region, below $-1.8 ~eV$, are now separated by $45.4 ~meV$ as shown in sub-figure (2).

{\par}It was never expected that this C$_{17}$Ge structure would display large SOC effect, still, the band structure exhibits ample SOC footprint in removing band degeneracies. The spin degeneracy removal, however small it be, can generate spin Hall current in devices and we focus on that in the next section. 

\vskip 1cm
\noindent
\textit{\textbf{Spin Hall Effect: }}
{\par}Intrinsic SOC is an interesting feature of materials as it generates an electric field within such substance, even in absence of any external perturbation. Due to the removal of band degeneracy in intrinsic spin-orbit coupled materials, this transverse electric field triggers spin current. This phenomena is known as spin Hall effect. For different semiconductors and metals this feature is prolific \cite{sinova2004}. The ab-initio calculations have produced excellent results but these type of direct calculations are excessively expensive as almost $10^6$ number of k-pints are needed for simple semiconducting systems like Ge, GaAs, etc. \cite{PhysRevLett.94.226601}. So, for larger systems with lesser symmetries direct ab-initio calculations become almost impossible. Qiao \etal successfully found an alternative way of calculation through Wannier interpolation and could reproduce the ab-initio results effectively \cite{qiao2018} and further applied on layered materials \cite{zhou2019}.

{\par}In presence of electric field along $\beta$ direction, the response of spin ($\sigma^{\gamma}$) current along the $\alpha$ direction can be calculated using linear response theory and the Kubo formula for spin Hall conductivity (SHC) tensor is given by:
\begin{equation}
\sigma_{\alpha\beta}^{\gamma} =
 -\frac{e^2}{\hbar V N_{\k}^3}\sum_{\k,n} f_{n\k} \Omega_{n,\alpha\beta}^{\gamma}(\k)
\end{equation}
where, $V$ is the volume of the unit cell, $N_{\k}$ is the number of k-points in Brilouin zone and $\Omega_{n,\alpha\beta}^{\gamma}(\k)$ is the Berry curvature like term expressed as:
\begin{equation} 
\Omega_{n,\alpha\beta}^{\gamma}(\k) = -\hbar^2 \sum_{m\ne n}
\frac{\textbf{Im}[\langle n\k|\{\hat{\sigma}_\gamma, \hat{v}_\alpha\}|m\k\rangle
	\langle m\k| \hat{v}_\beta|n\k\rangle ]}{(E_{n,\k}-E_{m,k})^2 - (\hbar \omega + \textbf{i} \eta)^2}
\end{equation}\label{SHC}
The term in curly bracket is coming from the spin current, $\hat{j}_{\alpha}^{\gamma}=\frac{1}{2}\{\frac{\hat{\sigma}_{\gamma}}{2},\hat{v}_{\alpha}\}$ and $\hat{v}_\alpha$ is the velocity operator. The matrix elements associated with these two quantities are calculated using Wannier interpolation techniques \cite{wannier90_pizzi2019}. In dc-SHC calculation, setting $\omega=0$ blows up the term in Eq.\ref{SHC}. To avoid that inter-band smearing has been introduced through an adaptive technique \cite{qiao2018,yates2007}.

\begin{figure}[h!]
	\centering
	\includegraphics[scale=0.3]{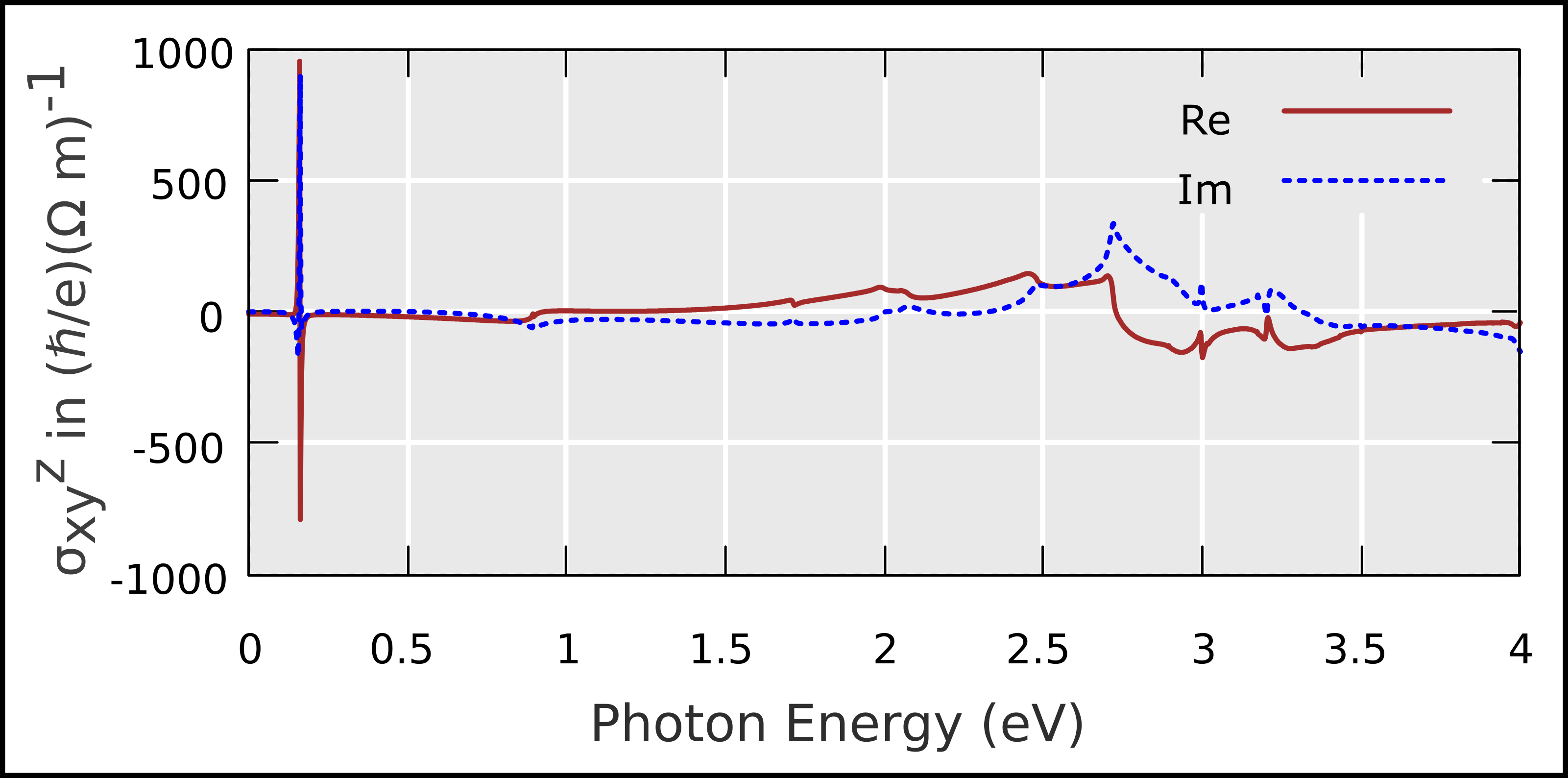}
	\caption{Variation of real and imaginary parts of ac-SHC $\sigma_{xy}^{z}$ with energy.}
	\label{fig4}
\end{figure}

{\par}For two dimensional systems, only two components of Hall conductivity tensor is important, $\sigma_{xy}^{z}$ and $\sigma_{yx}^{z}$. For symmetrical systems, like we are dealing here, $\sigma_{xy}^{z} = - \sigma_{yx}^{z}$. Though for dc-SHC, the imaginary part is always zero, for ac-SHC a variation of both real and imaginary parts of $\sigma_{xy}^z$ is expected.

{\par}In Fig. \ref{fig4}, we have plotted the variation of real and imaginary parts of ac spin Hall conductivity as a function of energy. As $\omega \rightarrow 0$, $\textbf{Im}[\sigma_{xy}^{z}] \rightarrow 0$ which is obvious for systems with band gap. The real part of SHC approaches a constant value of $-8.95 (\hbar/e)(\Omega m)^{-1}$ for zero energy, whereas, the imaginary part tends to zero. A sharp peak is located around the bandgap value $0.166 ~eV$ and after that they are almost flat till $0.75~eV$. Then there are random variations though not much sign change is observed. The real part changes sign $\sim 2.75 ~eV$ where a small peak of imaginary part appears. This peak is not as sharp as observed at $0.166 ~eV$. The highest value of $Re[\sigma_{xy}^z]$ is $ 956.61(\hbar/e)(\Omega m)^{-1}$ and $Im[\sigma_{xy}^z]$ is $908.14 (\hbar/e)(\Omega m)^{-1}$. These values are not as large as metallic or semimetallic systems \cite{zhou2019}, where SOC effect is much prominent. The C-$2p$ and Ge-$4p$ electrons are not expected to exhibit large spin Hall effect, however, this result is interesting enough to establish the effect of small doping of same group material in pristine graphene by displaying measurable SHC.

\subsection{Optical Properties:}
{\par}Optical properties of materials are directly related to their electronic structure. While for metallic systems both inter-band and intra-band transitions play role in optical properties, in semiconductors the optical property is mostly governed by inter-band jump. The absorption of a photon by any semiconducting system leads to the transition of an electron from VB to CB. So, the minimum amount of energy required to form the charge carriers should be greater than or equal to its band-gap. The absorption of a photon with energy similar to the bandgap produces a hole in the valence band and an electron in the conduction band. Photons with energy greater than bandgap initiates transition to higher level of conduction band. So, for optical property calculations always plenty of empty bands in CB are taken into consideration. In VB region, the number of electrons participating in the inter-band transitions can be calculated using the expression of effective electron number (n$_{eff}$) given as: 

\begin{equation}
n_{eff}(E_m) = \frac{2m}{Ne^2h^2}\int_0^{E_m}E . \epsilon^{(i)}_{\alpha\alpha}(E) dE;
\end{equation} 
{\par}Here, m, e and N are electron mass, charge and density. A saturation of the $n_{eff}$ with photon energy reflects the unavailability of electrons lower than that value for production of hole by jump towards CB.
Our calculated value has shown that electrons lying below $15 ~eV$ from VBM is not contributing in the optical transitions. So, all the plots are focused in the $0-15 ~eV$ range.

\begin{figure}[h]
	\centering{
		\includegraphics[scale=0.25]{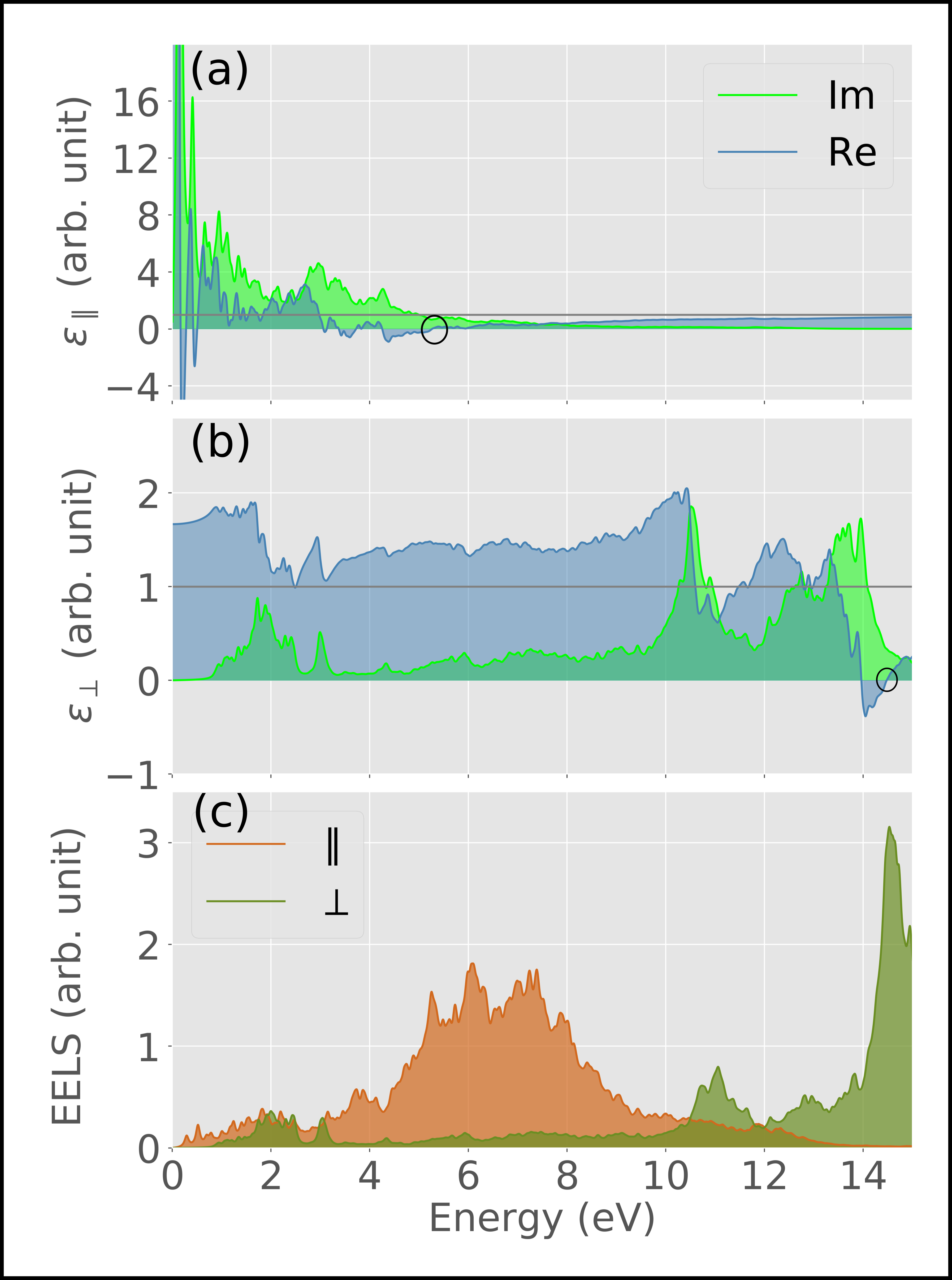} 
		\caption{Variation of real and imaginary part of diagonal element of dielectric tensor $\epsilon^{(r/i)}_{\alpha\beta}$ for (a) parallel polarization, (b) perpendicular polarization of electric field; (c) Electron energy loss spectrum (EELS). The plasma frequency is shown by circular pointer.}
		\label{fig5}
	}
\end{figure}

{\par}In Fig.\ref{fig5}, we have depicted the real and imaginary parts of diagonal element of dielectric tensor $\epsilon^{(r/i)}_{\alpha\beta}$ and electron energy loss spectrum (EELS) as a function of energy. There are two significantly different polarization of electric fields to be considered, one is parallel to the plane of the 2D material (denoted by $\epsilon_\parallel$) and perpendicular to that plane (denoted by $\epsilon_\perp$). The two orthogonal directions (say, x and y) in the plane are symmetric, so, $\epsilon_\parallel=\epsilon_{xx}=\epsilon_{yy}$.

{\par} Both the real and imaginary parts of parallel dielectric constants $\epsilon^{(r/i)}_\parallel$ show rapid fluctuations in lower energy region. In addition, $\epsilon^{(i)}_\parallel$ changes sign four times in this energy region. $\epsilon^{(r/i)}_\parallel$ becomes almost flat only at high energy region ($>8 ~eV$).

{\par} The real part of dielectric constant crosses the zero axis from negative to positive at five different energies, two below $1eV$ and three in the $3-6~eV$ range. The plasma frequency is defined as the frequency when $\epsilon^{(r)}$ shows positive slope while crossing the zero axis and $\epsilon^{(i)} \langle 1$. The fourth zero crossing energy $5.267 ~eV$ satisfies both the conditions and identified as the plasma frequency (encircled in Fig.\ref{fig5}) for parallelly polarized electric field. This can be readily verified by the peak in EELS. In most of the semiconductors the plasma frequency lies deep into ultraviolet region and germagraphene is not an exception.

{\par}  We find relatively flat $\epsilon^{(r/i)}_\perp$ than its perpendicular counterpart. The values are smaller as well. The imaginary part shows a sharp peak around $\sim 10.5~eV$. Unlike $\epsilon^{(r)}_\parallel$, $\epsilon^{(r)}_\perp$ becomes negative to positive only once, at $14.48~eV$ which is the plasma frequency for perpendicular polarization.
  
{\par}Electron energy loss spectrum (EELS) is a properties which can be verified through experiments. In  Fig.\ref{fig5}(c) the plot of EELS is presented. While the EELS for parallel polarization takes a bell shape centring around $6.5~eV$, the perpendicular counterpart exhibit significant growth only after $10~eV$. There is a crossover of these two at $\sim 10.3~eV$ and thereafter, the perpendicular part is dominant. The sharp peak of EELS$_\perp$ is at the corresponding plasma frequency $14.48~eV$ as found from dielectric constant plot.

\vskip 1cm
\noindent
\textit{\textbf{Absorption coefficient, refractive index and  optical conductivity: }}
{\par}Transition of electron from VB to CB is a result of absorption of photonic energy. So, absorption coefficient preserves the signature of not only band gap but also the full characteristics of bands. Near the Fermi energy, such correspondence can be readily established.

\begin{figure}[t]
	\centering{
		\includegraphics[scale=0.25]{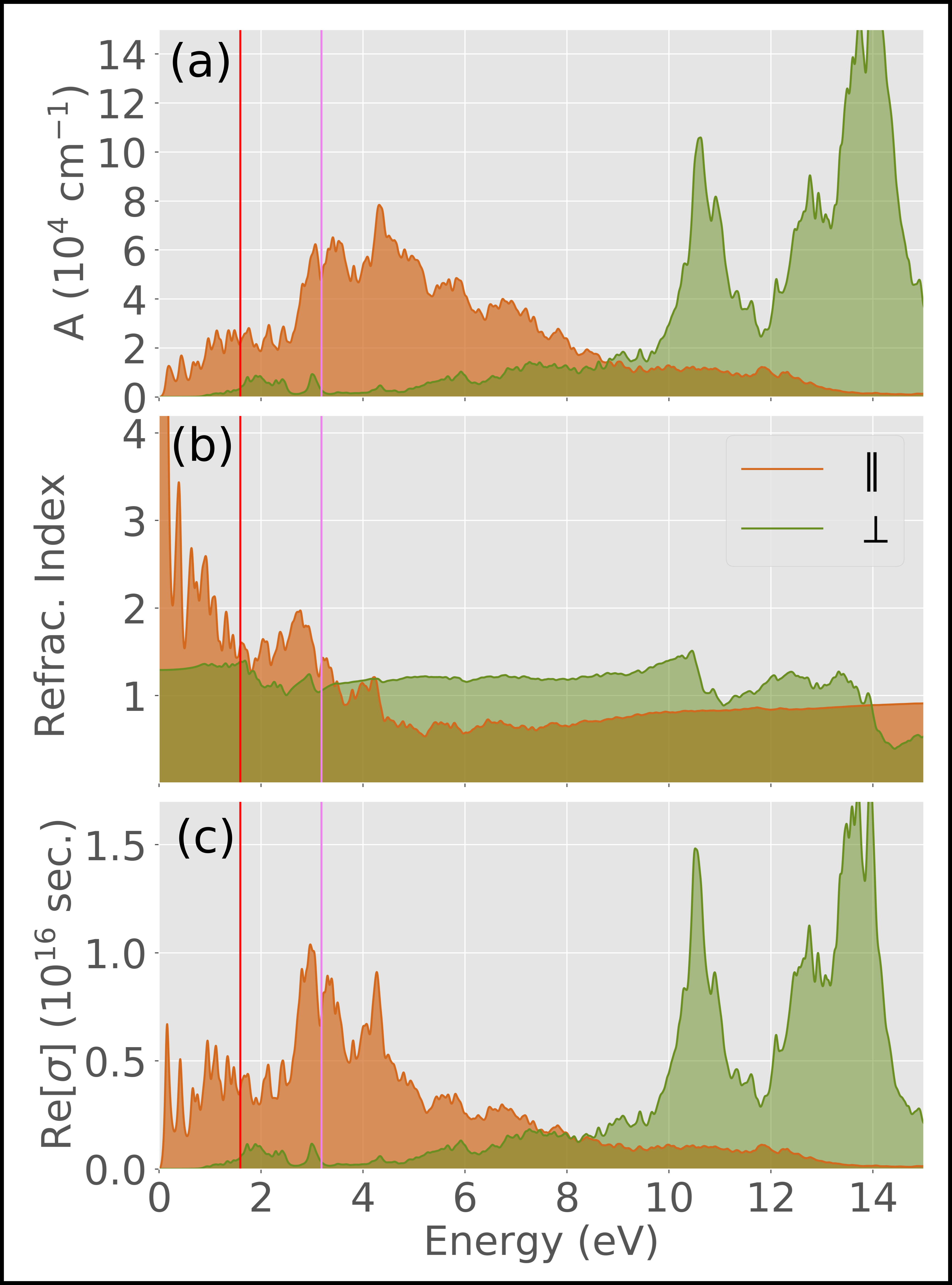} 
		\caption{ Graphical (a) absorption spectra, (b) refractive index, and (c) optical conductivity  as a function of photon energy for parallel and perpendicular polarizations of electric field.}  
		\label{fig6}
	}
\end{figure}

{\par}In Fig.~\ref{fig6}(a) $\&$ (b), we plot absorption coefficient as a function of photon energy.
To study the one to one correspondence between the electronic structure and photonic absorption we have to look back to Fig.\ref{fig2}.
The very first small peak in the absorption coefficient (A) for parallel polarization is at $\sim 0.17 ~eV$ corresponds to the band gap of the material. The second peak $\sim 0.45 ~eV$ is between VBM and the small hump of LU band in $\Gamma-K$ direction, the third peak around $0.90 ~eV$ is between the flat region of HO$-1$ around $\Gamma$ and CBM. In visible region, the absorption coefficient is showing a gradual increase first, then at around $2.6 ~eV$ starts to rise stiffly. The jump from around $2.6 ~eV$ to $3.1 ~eV$ is attributed from HO to LU$+2$ transitions as well as from the HO$-1$ to LU$+1$ transitions around the midway of $\Gamma-K$ and $\Gamma-M$ path. This region corresponds to blue-violet colour region. Blue to ultraviolet light absorption is particularly useful for safety of different optical devices through coating by those high absorbing materials. 

{\par} The absorption coefficient for perpendicular polarization ($A_\perp$) is negligible with respect to the parallel counterpart till $8~eV$. After that it is dominant as observed for dielectric constant. 

{\par}In the application for coating on devices another important optical property is the refractive index.
The real refractive index of germagraphene structure is depicted as function of photon energy in Fig.~\ref{fig6}(b). In the visible range ($1.55-3.0 ~eV$), the refractive index for parallel polarized electric field varies from $1.27$ to $1.95$ having an average $\approx1.64$.  For comparison, we quote the refractive index of flint glass as $1.60-1.62$, so, this germagraphene material has similar average refractive index to glass for parallel electric field. For perpendicular field, the variation is in $1.0-1.4$ range.

{\par} There is a direct connection of optical absorption spectra with the imaginary part of the refractive index through the Eq.\ref{abs}. Therefore, the nature of refractive index plots are expected to replicate the peaks and valleys of absorption spectra.
 
{\par}The optical conductivity in Fig.~\ref{fig6}(c) shows similar character to absorption coefficient which is logically evident. As more electrons move to CB through photon absorption, the carrier concentration increases resulting in higher optical conductivity. The signature of semiconductors, a gap at the beginning of optical conductivity is clearly visible. The first peak is around the band-gap value of $0.166 ~eV$. The peak is sharper than that of the absorption coefficient because the absorption of one photon creates two carriers, one electron and one hole.

\subsection{Thermal Properties:}
{\par}In materials electrons carry both charge and heat. The temperature gradient of any material produces an electric field which opposes the natural diffusion of electrons. So, the electronic conductivity becomes dependent on temperature. The opposing electric field produces the so called Seebeck voltage. The Seebeck coefficient and electronic conductivity depends on this Seebeck voltage. As a gradient dependent quantity, the Seebeck voltage is directly correlated with the grade of asymmetry of electronic distribution around Fermi energy.

{\par}A narrow band gap is one of the primary requirement to identify thermoelectric materials for further extensive study, high carrier mobility (low effective mass) is another. Low bandgap helps to pump electrons for conduction even with the help of small thermal excitation with tendency of increasing the electrical conductivity. Besides, direct nature of gap is an additive benefit. Further, low effective mass ensures the high mobility of the carriers, thus, increases the electrical conductivity. However, the decisive factors for searching a promising thermoelectric material consist of a variety of different interconnected properties. From electronic point of view, electrical conductivity ($\sigma$), Seebeck coefficient (S) or the power factor ($\sigma S^2$) are the key parameters to dictate the thermoelectric performance.

\begin{figure}[t!]
	\centering
	\includegraphics[scale=0.25]{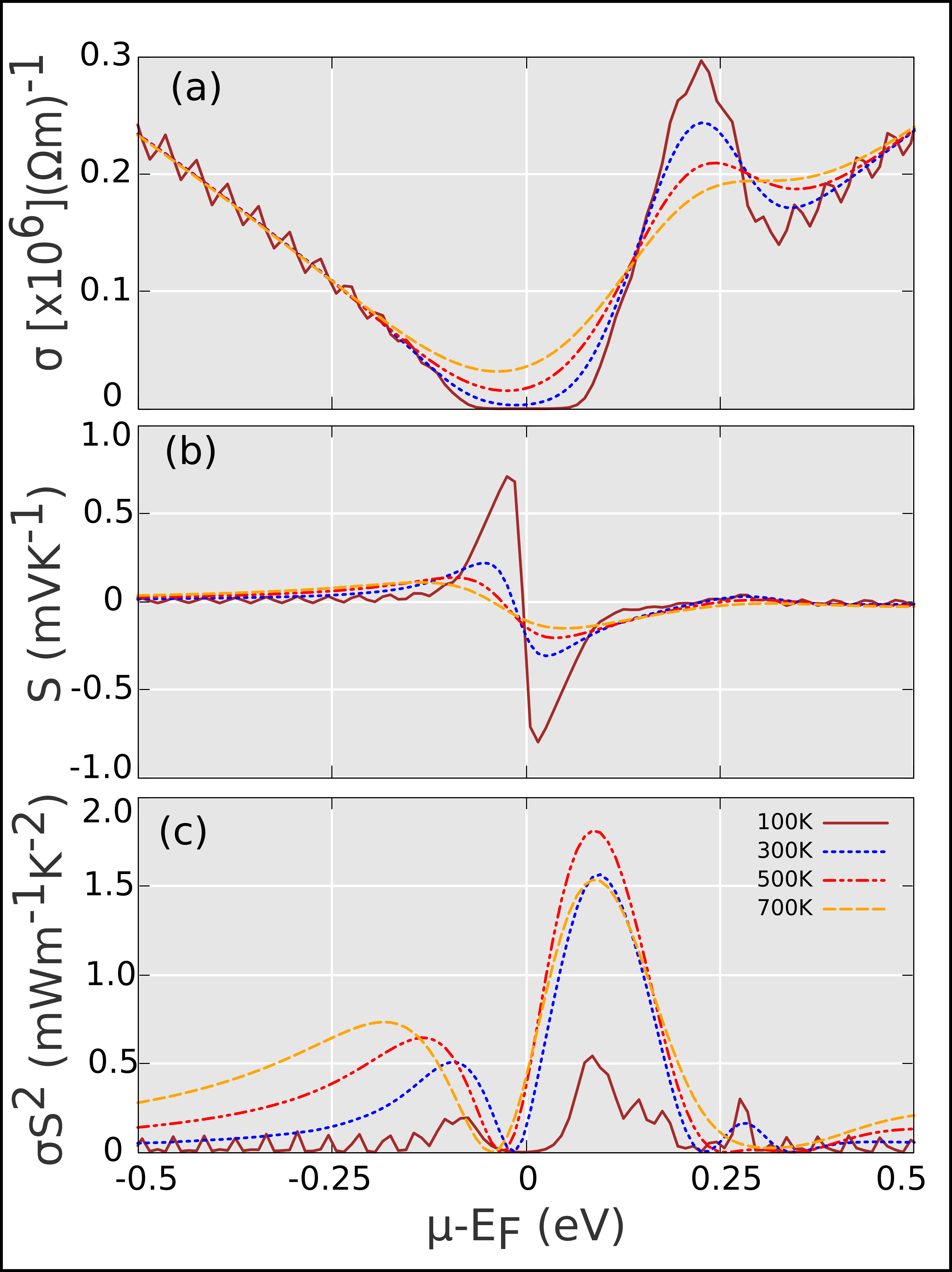}
	\caption{ Thermal properties of C$_{17}$Ge: (a) Electrical conductivity ($\sigma$), (b) Seebeck Coefficient (S)  and (c) power factor ($\sigma S^2$) as a function of chemical potential.}
	\label{fig7}
\end{figure}

{\par}In Fig.\ref{fig7} we have plotted electrical conductivity, Seebeck coefficient and power factor at different temperature (100, 300, 500, 700 K) for C$_{17}$Ge as functions of chemical potential ($\mu$) as measured from $E_F$. The zero of chemical potential represents the pristine condition and  positive (negative) value indicates the electron (hole) doping. In pristine condition, the conductivity rises with temperature which is a sign of the semiconducting nature of C$_{17}$Ge.

{\par} We have seen in Fig.\ref{fig2} that the DOS has much higher peak around CBM than around VBM. This means more electron concentration in CB edge than hole concentration at VB edge. Hence, asymmetric plot about the zero chemical potential is justified and more electrical conductivity at n-side is a result of higher carrier concentration near CBM.

{\par}The electrical conductivity for $100K$ has sharper variations with comparison to the other temperature plots. Conductivity at room temperature ($300K$) rises gradually, attains its highest value $0.24\times 10^6 (\Omega m)^{-1}$ at $\mu = 0.226$ and then shows a dip at $\sim \mu=0.31$ before staring to rise again. Similar variation is seen for $500K$, however at $700K$ this dip is absent.

{\par}  At room temperature (300 K) the highest value of Seebeck coefficient is $ -309 ~\mu V/K$ at $\mu=0.0256~eV$ for n-type doping and $ 220 ~\mu V/K$ at $\mu=-0.0544~eV$ for p-type doping. These values are much higher than the value for pristine graphene, graphene based heterostructure \cite{seebeck,duan2016}, and comparable to Bi$_2$Te$_3$ \cite{cheng2014}. A comparison with other thermoelectric materials is presented in Table. \ref{tab1} \cite{wang2017, lee2015, cheng2014}. High Seebeck effect can be mapped with the rapid variance of DOS. High energy gradient of DOS ($\frac{d DOS}{d E}$) is always beneficial for thermoelectric performance\cite{wu2018}. Under doping, if the Fermi level moves to high DOS peak region, a large asymmetry of thermoelectric power is developed between the hot electrons of energy higher than $E_F$ and cold electrons of energy below Fermi level. Higher anisotropy means higher S, as we can verify for our sample, where the sharp peaks in CB (near CBM) is responsible for highest peaks in Seebeck coefficients. Lower anisotropy of DOS in VB is the reason of lower peak values in $\mu < 0$ region than $\mu > 0$.  As temperature rises, the increase in Joule heating brings down the value of Seebeck coefficient. So, in the figure, we see less Seebeck coefficient for higher temperature.

{\par} The power factor of any material is defined as $PF=\sigma S^2$. To understand the electronic structural effect on power factor we have to look back to the carrier effective masses. Higher effective mass is favourable for Seebeck coefficient, whereas it has a negative effect on electrical conductivity. In PF, the second order dependence on S overwhelms the unfavourable effect on $\sigma$. C$_{17}$Ge has an unusual semiconductor with a very flat LU and very sharp HO band. This in turn initiates the large anisotropy between hole and electron effective masses. So, the power factor follows that anisotropy and for positive $\mu$ the PF peaks are significantly higher than negative $\mu$. Even the peak for $300K$ outplays the peak of $700K$ for n-type doping. For $\mu<0$, higher temperature possesses higher maxima.

{\par} The power factor at $300 K$ is maximum ($1.56 ~mW m^{-1} K^{-2}$) near $\mu = 0.096~eV$ due to significant increase of the electrical conductivity at the high electron doping range. The value of power factor is in between two orthodox thermoelectric materials, Bi$_2$Te$_3$ and Sb$_2$Te$_3$ \cite{cheng2014,lee2015} and better than some other proposed materials \cite{wang2012, guilmeau2015}. Moreover, the excellent value of S establishes germagraphene as a potential candidate for thermoelectric application.

\begin{table*}[]
	\centering
	\begin{tabular}{c|cc|c|c|c}
		\hline
		& \multicolumn{2}{c|}{\textbf{Effective Mass ($m_0$)}} &  &  &  \\ \cline{2-3}
		\multirow{-2}{*}{\textbf{Material}} & \textbf{Hole} & \textbf{Electron} & \multirow{-2}{*}{\textbf{\begin{tabular}[c]{@{}c@{}}Seebeck Coefficient\\ ($ \mu V/K$)\end{tabular}}} & \multirow{-2}{*}{\textbf{\begin{tabular}[c]{@{}c@{}}Power Factor\\ ($mW m^{-1} K^{-2}$)\end{tabular}}} & \multirow{-2}{*}{\textbf{\begin{tabular}[c]{@{}c@{}}Work Function\\ (eV)\end{tabular}}} \\ \hline
		\rowcolor[HTML]{EFEFEF} 
		\textbf{C$_{17}$Ge} & -0.00292 & \begin{tabular}[c]{@{}c@{}}heavy:0.59528\\ light:0.00291\end{tabular} & 309 & 1.56 & 4.361 \\
		\textbf{Graphene} & -- & -- & 31 \cite{seebeck} & $\sim 0$ & \cellcolor[HTML]{EFEFEF}4.390 \\
		\textbf{Germanene} & -- & -- & -- & -- & \cellcolor[HTML]{EFEFEF}4.682 \\
		\textbf{GeAs$_2$}\cite{wang2017} & -0.20 & 0.15 & 400 & 0.55 & -- \\
		\textbf{MoS$_2$} \cite{rasmussen2015} & -0.56 & 0.55 & 500-550 \cite{ghosh2015} & 8.50\cite{hippalgaonkar2017} & 6.32 \\
		\textbf{WSe$_2$} \cite{rasmussen2015} & -0.42 & 0.46 & 450-500 \cite{ghosh2015} & 3.20 \cite{yoshida2016} & 5.60 \\ \hline
		\textbf{Bi$_2$Te$_3$} \cite{yavorsky2011} & -0.024 & 0.178 & 323.0  & $\sim$2.60\cite{cheng2014}  &  --\\
		\textbf{Sb$_2$Te$_3$}\cite{yavorsky2011} & -0.054 & 0.045 & 115.6  & $\sim$ 1.00 \cite{lee2015}  &  4.30\\
	\hline
	\end{tabular}
	\caption{Effective mass, Seebeck coefficient, power factor and work function of C$_{17}$Ge at $300K$ compared with other 2D (graphene, germanene, MoS$_2$, GeAs$_2$, WSe$_2$) and 3D (Bi$_2$Te$_3$, Sb$_2$Te$_3$) materials. The results of present work are in shaded cells.}
	\label{tab1}
\end{table*}

\subsection{Work function:}  
{\par} The work-function ($\Phi$) of any material is defined as the energy required to remove an electron from the surface of the material. The highest energy of electron in material is its Fermi energy ($E_F$), and the energy at the point just outside the surface is originally the vacuum energy ($E_{vacuum}$). Work function is the difference of these two:
\begin{equation}
	\Phi = E_{vacuum} - E_F
\end{equation}\label{eq_WF}
In intrinsic semiconductors $E_F$ is the mid point energy of VBM and CBM, so, in absolute scale, the orientation of the bands can be found accordingly.

{\par} It is well-known that the PBE functional is not adequate enough to determine the work function correctly, whereas, Heyd$-$Scuseria$-$Ernzerhof (HSE) \cite{HSEtheo, HSEtheo1} screened hybrid functional has shown great success for semiconductors. The failure of Local Density Approximation (LDA) and the PBE based Generalized Gradient Approximation (PBE-GGA) in determining the band gap exactly is also evident \cite{pplb}. HSE functional can solve that problem in most of the cases. 

{\par} We have done the HSE calculation for the same optimized germagraphene unit cell and same energy and charge density cut-off. The k and q point grids are taken as $4\times4\times1$ and Optimized Norm-Conserving Vanderbilt (ONCV) pseudopotentials are used. The HSE calculated band-gap is $0.237 ~eV$ which is $\approx 43\%$ higher than PBE calculated value ($0.166 ~eV$).

\begin{figure}[h!]
	\centering
	\includegraphics[scale=0.3]{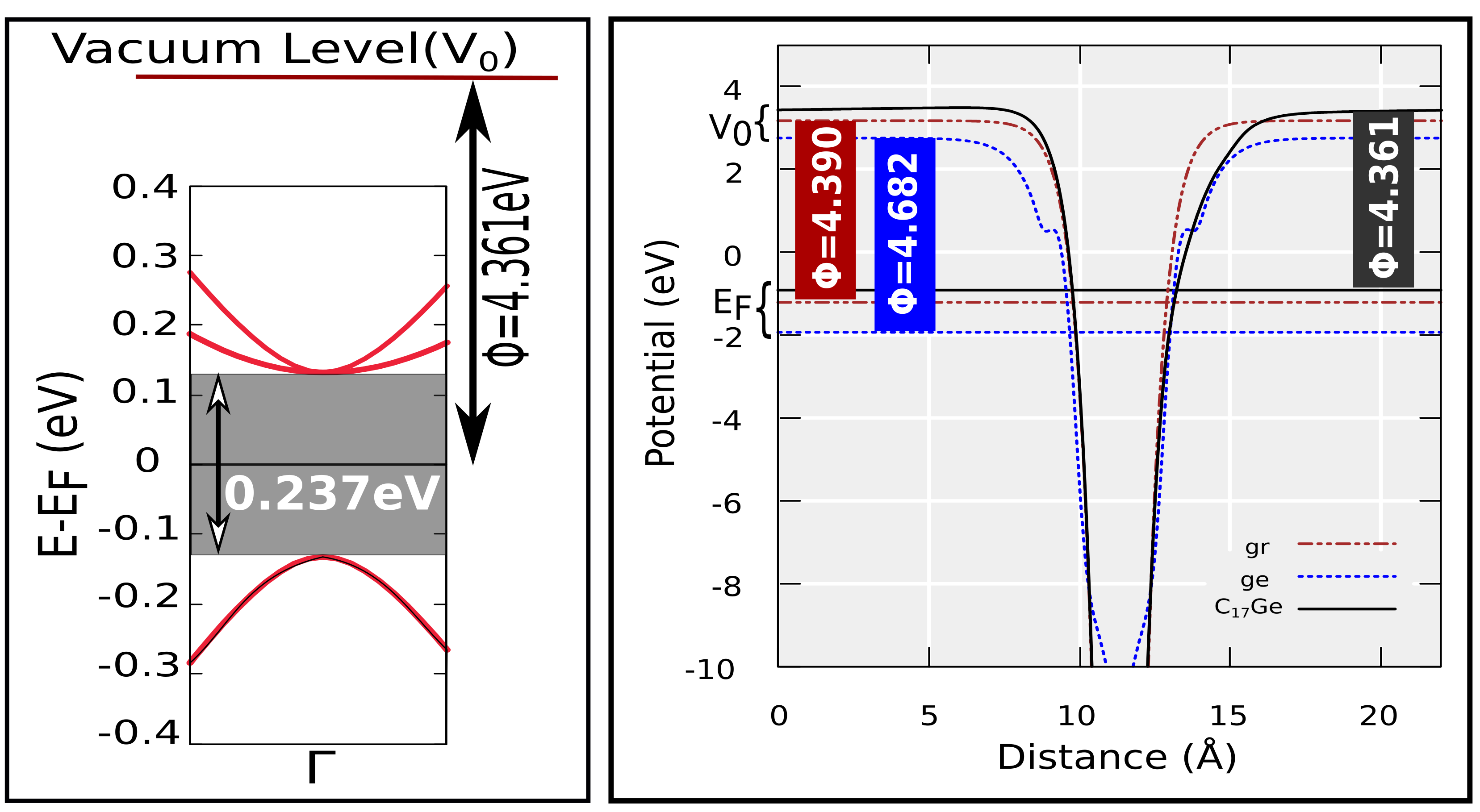}
	\caption{ (left) HSE calculated work function and band-gap of C$_{17}$Ge. (right) Electrostatic potential plot for graphene (red), germanene (blue) and C$_{17}$Ge (black) for calculation of corresponding work functions ($\Phi$)}
	\label{fig8}
	\end{figure}

{\par} In Fig.\ref{fig8} we plot the electrostatic potentials of germagraphene in comparison with graphene and germanene. The vacuum level is different for different systems and dependent on the software packages, so, for comparison this vacuum level is always standardized to the zero energy. Using Eq.\ref{eq_WF} the work functions of these three materials are calculated. Our predicted value of graphene monolayer ($4.39 ~\AA$) \cite{sujoyC3N4} is at par with experimental result $4.56~eV$ \cite{yan2012} and better than the PBE estimated value $\approx 4.23 ~eV$ \cite{ziegler2011}. The $\Phi$ of germanene is $4.682 ~eV$ similar to the already reported value of $4.688~eV$ \cite{li2016}. The HSE calculated work function of C$_{17}$Ge is $4.361 ~eV$ which is lower than both of these. Lower work function is beneficial for removal of electron from material surface. The $\Phi$ is of particular interest in the field of photo-catalytic devices through heterojunction formation. Being a material of lower work function than graphene, it may be considered as an alternative to those heterojunction devices where graphene is used now.

\section{Conclusion}
{\par} We show that the stable buckled C$_{17}$Ge structure is a direct band gap semiconductor having a PBE calculated band gap of $0.166 ~eV$. The band degeneracy at the conduction band minima is left when spin orbit coupling is considered. This spin dependent electronic property leaves its footprint on spin hall conduction with highest value of conductivity as $956.61 (\hbar/e)(\Omega m)^{-1}$. Though this value is not large but effective enough to showcase the effect of anisotropy induced by Ge on the spin dependent electronic property. We also find an exceptionally small hole and light-electron effective mass. These are preferable for optoelectronic devices like infrared detector, photovoltaic devices etc. where high mobility is a perquisite criteria.

{\par} The optical and thermal properties of C$_{17}$Ge are also been explored. This material has exhibited significant blue to ultraviolet light absorption. So, we predict if devices are coated by this material then they can avoid the harm from lower UV region. The refractive index comparable to flint glass is an additional advantage of this material in optoelectronic world.

{\par} The narrow band gap of germagraphene makes it a potential candidate for thermoelectric applications. C$_{17}$Ge has shown relatively larger value of Seebeck coefficient and power factor comparable to traditional thermoelectric materials such as Bi$_2$Te$_3$ and Sb$_2$Te$_3$ at room temperature. Thus, C$_{17}$Ge affirms brilliant thermoelectric performance. In a sum up, we can conclude that the efficiency of germagraphene in futuristic applications is quite promising. 

\section*{Acknowledgement}
The authors acknowledge the contribution of \textit{\textbf {Late Prof. Abhijit Mookerjee}} in forming the problem at the initial stage. SD wants to thank \textit{\textbf {Dr. Prashant Singh}} of Ames Lab., USA and \textit{\textbf {Dr. Chhanda Basu Chaudhuri}} of Lady Brabourne College, India for their suggestions.

\bibliography{germagraphene}
\bibliographystyle{rsc}

\end{document}